# Hybrid Lattice Boltzmann/Dynamic Self-Consistent Field Simulations of Microphase Separation and Vesicle Formation in Block Copolymer Systems

*Liangshun Zhang[1]\**, *Agur Sevink[2]+*, *and Friederike Schmid[3]\**

\*Institute of physics, Johannes Gutenberg Universität Mainz, D-55099 Mainz, Germany
+Leiden Institute of Chemistry, Universiteit Leiden, 2300 RA Leiden, Niederlande

We present a hybrid numerical method to introduce hydrodynamics in dynamic self-consistent field (SCF) studies of inhomogeneous polymer systems. It solves a set of coupled dynamical equations: The Navier-Stokes equations for the fluid flow, and SCF-based convection-diffusion equations for the evolution of the local monomer compositions. The Navier-Stokes equations are simulated by the lattice Boltzmann method and the dynamic self-consistent field equations are solved by a finite difference scheme. Two applications are presented: First, we study microphase separation in symmetric and asymmetric block copolymer melts with various values of shear and bulk viscosities, comparing the results to those obtained with purely diffusive dynamics. Second, we investigate the effect of hydrodynamics on vesicle formation in amphiphilic block copolymer solutions. In agreement with previous studies, hydrodynamic interactions are found to have little effect on the microphase separation at early times, but they substantially accelerate the process of structure formation at later times. Furthermore, they also contribute to selecting the pathway of vesicle formation, favoring spherical intermediates over aspherical (disklike) ones.

## 1. INTRODUCTION

Block copolymers, where chemically distinct building blocks are connected by covalent bonds, can spontaneously self-assemble into a variety of nanoscale ordered microstructures. For example, melts of block copolymers may organize themselves as lamellae, cylinders, spheres, or complex bicontinuous structures.[1,2] Amphiphilic block copolymers in solution[3,4] may self-assemble into, e. g., spherical and cylindrical micelles, bilayer membranes, and vesicles. These microstructures can be tailored by varying the copolymer composition and segregation degree between the two blocks and have attracted widespread interests in both fundamental and technological contexts.

During the formation of ordered microstructures following a quench, the long-range hydrodynamic interactions strongly influence the domain coarsening.[5] This can already be observed in simple binary mixture.[6,7,8,9] Theories as well as numerical simulations of spinodal decomposition of binary mixture show that domains grow according to a power law behaviour and the growth exponent is a universal constant, which is different for systems with and without hydrodynamic interactions. For more complex systems, the effect of hydrodynamic effects on phase ordering and coarsening is less clear. A general feature seems to be that hydrodynamic flows speed up the coarsening process and help to prevent trapping in metastable states.

In order to obtain insight in these phenomena, different simulation models and methods have been proposed.[10,11,12,13,14,15,16,17] For instance, Yokojima and Shiwa used cell-dynamics to investigate the hydrodynamic coarsening of microstructures for quenched symmetric and asymmetric block copolymers.[11] The hydrodynamic effects are introduced by a velocity field, which is defined in terms of a vertical vorticity. The calculations indicate that the hydrodynamic flows are effective in the ordering process of symmetric block copolymers, while they are not important for the asymmetric block copolymers. Going beyond cell dynamics, the Lattice Boltzmann (LB) method has emerged as a powerful tool to solve the Navier-Stokes equations on a lattice.[18,19] Phase separation and interface dynamics can be incorporated into the LB method either directly by introducing intermolecular interactions, or via free energy functional based concepts, or using other mesoscopic equations.[12,20,21,22,23,24] The LB method is computationally efficient and comparatively easy to parallelize for large size simulations. Gonnell, Xu and co-workers have combined the LB method with a time-dependent Ginzburg-Landau model to study microphase separation in block copolymers.[12,13,14,15] The approach deals with hydrodynamic interaction, but does not explicitly take into account the conformational entropy of polymer chains.

---

[1] E-mail: shunshun122@gmail.com
[2] Email: a.sevink@chem.leidenuniv.nl
[3] Corresponding Author: E-Mail: Friederike.Schmid@Uni-Mainz.DE

On the other hand, polymer-specific dynamic models like the dynamic self-consistent field (SCF) theory are becoming increasingly popular for studying kinetic processes in inhomogeneous polymer systems.[25,26,27,28,29] Unlike phenomenological Ginzburg-Landau theories, the SCF theory accounts for the chain connectivity and provides a unified treatment of polymer systems in intermediate and strong segregation regions.[30,31] However, the dynamics in most dynamic SCF simulation methods is purely diffusive and does not account for hydrodynamic effects. A few efforts have been made in recent years to overcome this drawback. Maurits et al. have devised a hydrodynamic SCF approach which combines the dynamic SCF equations with a simple hydrodynamic model based on the Darcy's law for permeable media, i.e., hydrodynamic interactions were assumed to be purely local. This approach was used to examine the influence of hydrodynamic flows on the evolution process of microstructures.[32] Already at this level, the convective flows were found to accelerate domain growth and the removal of defects in the late stage of ordering. Honda and Kawakatsu combined dynamic SCF theory with the Navier-Stokes equations to investigate the hydrodynamic effects on the disorder-to-order transition of block copolymers.[33] A quasistatic approximation was developed to solve the Navier-Stokes equations. They found that a system with lower viscosity has a lower free energy density. Song et al. combined a dynamic SCF approach with the lattice Boltzmann method to study the effects of hydrodynamic interactions on the microphase separation process of block copolymers and defect evolution of lamellar microstructures in two-dimensional space.[23,24] In their method, the convection-diffusion equation for the evolution of the order parameter is numerically solved by the lattice Boltzmann method, which may generate unphysical flows due to spurious terms.[34] Hall et al. have developed a complex framework named hydrodynamic self-consistent field theory (HSCF)[35,36] for dense melts with arbitrary constitutive equations, which they solve using a pseudo-spectral method.

In the present work, we propose a new hybrid approach to implementing hydrodynamic interactions in dynamic SCF simulations. Starting from a minimal model of coupled dynamic SCF and Navier-Stokes equations, we propose a hybrid method to simulate them efficiently. The dynamic SCF equations are solved by a finite difference scheme,[14,37,38] which reduces the spurious current generation. The continuity equation and the Navier-Stokes equations are solved by the multiple-relaxation-time lattice Boltzmann method,[39,40,41] which reduces the danger of numerical instabilities. The method can be applied both to polymer melts and solutions. Here we demonstrate its potential by studying two different types of systems. As a first test case, we investigate the kinetics of microphase separation in block copolymer melts. Then, we study the influence of hydrodynamics on vesicle formation in solutions of amphiphilic block copolymers.

This second application is motivated by the high current interest in vesicle nanostructures. Vesicles are closed membranes with usually spherical topology. They are biologically important as model systems for plasma membranes and sub-cellular compartments in the living cells. Because of the analogy to the natural systems and the compositional simplicity, vesicles also serve as ideal platform for understanding cellular membrane remodelling processes such as fusion and fission of cells.[42,43,44,45] Apart from their vital importance in biology, vesicles also have extraordinary potential applications in nanotechnology and biotechnology,[46,47,48,49] for instance as microreactors or as nanocompartments for the confinement, transport, and manipulation of chemical cargo.

Although extensive experimental and theoretical studies have been carried out in order to understand the self-assembly of amphiphilic molecules into vesicles and the remodelling of vesicles,[50,51], much less is known about the kinetics of vesicle formation and vesicle remodeling under non-equilibrium conditions. The mechanisms for such processes are still subject to speculation.[52] Computer simulations provide a valuable alternative tool for studying such systems at the microscopic and mescoscopic levels.[53,54,55] Various coarse-grained particle-based molecular simulation approaches have been applied to investigate evolution and remodeling processes of vesicles.[56,57,58,59,60,61,62,63,64] Among these, those using implicit solvent models are particularly efficient, but incorporating hydrodynamics is difficult. On the other hand, most studies based on models with explicit solvent were aimed at detailed simulations of phenomena on small time and length scales, or focussed on flat bilayers to study specific aspects of membrane remodelling, due to the high computational costs of simulating such fully explicit models. Field-theoretical methods provide a promising alternative approach to simulating systems with larger volume over a longer time period. A number of recent SCF and dynamic SCF studies have therefore been devoted to studying membrane dynamics and fusion, and vesicle formation and fusion.[65,66,67,68,69,70,71,72] Sevink and Zvelindovsky used dynamic SCF simulations to investigate the self-assembly behavior of block copolymer droplets in solution with two hydrophobic blocks of different hydrophobic strength.[67,68] He and Schmid adopted a dynamic SCF method to examine the spontaneous vesicle formation in dilute solution of amphiphilic block copolymers.[69,70,71] Spherical and cylindrical micelles, toroidal and net-cage micelles as well as vesicles could be observed by tailoring the concentration and the hydrophilicity of the individual blocks. Moreover, a new pathway of vesicle formation was identified. These results are encouraging. However, hydrodynamic interactions, which should play an important role in solution, were neglected. Thus the effects of hydrodynamic flows on vesicle formation and vesicle remodeling are still not clear. This motivates us to reconsider the problem using our new hybrid simulation scheme.

The remainder of the paper is organized as follows: In the next section, we present the model equations and the hybrid simulation method. Then we discuss two applications in section 3. We first apply the method to studying symmetric and asymmetric block copolymer melts (section 3.1). We recover the previous results cited above and analyze the role of bulk and shear viscosity. Then, we turn to studying vesicle formation in solution (section 3.2). The hydrodynamic flows are found to accelerate the vesicle formation and vesicle fusion in the amphiphilic block copolymer/solvent system significantly. The intermediate processes of vesicle formation and vesicle fusion are presented in detail. We summarize and conclude in section 4.

## 2. MODEL EQUATIONS AND HYBRID SIMULATION METHOD

We consider a system with volume $V$, containing $n_p$ block copolymers (**AB**) and $n_s$ solvent beads (**S**). (In copolymer melts, we simply set $n_s$ to zero.) All copolymer molecules are represented by Gaussian chains $\mathbf{A}_n\mathbf{B}_m$, consisting of $N_p=n+m$ beads with relative A fraction $f=n/N_p$ and statistical segment length $a$. The solvent consists of single "beads" ($N_s=1$), each representing a collective of solvent molecules. The volume and mass of beads are denoted by $v_I$ and $m_I$, respectively, for beads of type $I=A,B,$ or $S$.

In our SCF theory, the free energy functional of the amphiphilic block copolymer/solvent system is defined as follows[26,73]

$$F\left[\{\rho_I\}\right] = -k_BT \ln \frac{\Phi_p^{n_p}\Phi_s^{n_s}}{n_p!n_s!} - \sum_I \int d\mathbf{r}\rho_I(\mathbf{r})U_I(\mathbf{r}) \\ + \frac{1}{2}\sum_{I,J}\int d\mathbf{r}d\mathbf{r}'\varepsilon_{IJ}(|\mathbf{r}-\mathbf{r}'|)\rho_I(\mathbf{r})\rho_J(\mathbf{r}') \\ + \frac{\kappa_H}{2}\int d\mathbf{r}\left(\sum_I v_I\,\rho_I - 1\right)^2 \quad (1)$$

Here, $k_B$ is the Boltzmann's constant, $T$ is the temperature, the subscripts $I$ and $J \in \{A, B, S\}$ are component indices, $\Phi_P$ is the intramolecular partition function for ideal Gaussian copolymer chains in the external fields $U_A$ and $U_B$, $\Phi_S$ is the partition function of solvent beads subject to the external field $U_S$, and $\rho_A$, $\rho_B$, and $\rho_S$ are the local number densities of A, B, and S beads, respectively. The third term on the right hand side of Eq. (1) is the contribution from the non-ideal mean-field interactions. It is characterized by the energetic interaction $\varepsilon_{IJ}(|\mathbf{r}-\mathbf{r}'|)$ between beads $I$ and $J$ at positions $\mathbf{r}$ and $\mathbf{r}'$, which has the kernel $\varepsilon_{IJ}(r) = \varepsilon_{IJ}^0 \left(\frac{3}{2\pi a^2}\right)^{\frac{3}{2}} \exp\left(-\frac{3r^2}{2a^2}\right)$ with constant cohesive interaction parameter $\varepsilon_{IJ}^0$. The last term of the free energy functional is a Helfand penalty that ensures constant local density by means of the Helfand compressibility parameter $\kappa_H$. The potentials $U_I$ are conjugated to the local density fields $\rho_I$ and must be determined self-consistently from the relation

$$\rho_A[\{U_I\}](\mathbf{r}) = \frac{n_p}{\Phi_p} \sum_{s=1}^{n} \int_{V^{N_p}} d\mathbf{R}_1 \cdots d\mathbf{R}_{N_p} \delta(\mathbf{r} - \mathbf{R}_s)$$
$$\times \exp\left(-\frac{1}{k_B T}\left(H^G + \sum_{s'=1}^{N_p} U_{I_{s'}}(\mathbf{R}_{s'})\right)\right)$$

$$\rho_B[\{U_I\}](\mathbf{r}) = \frac{n_p}{\Phi_p} \sum_{s=n+1}^{N_p} \int_{V^{N_p}} d\mathbf{R}_1 \cdots d\mathbf{R}_{N_p} \delta(\mathbf{r} - \mathbf{R}_s) \qquad (2)$$
$$\times \exp\left(-\frac{1}{k_B T}\left(H^G + \sum_{s'=1}^{N_p} U_{I_{s'}}(\mathbf{R}_{s'})\right)\right)$$

$$\rho_S[\{U_I\}](\mathbf{r}) = \frac{n_s}{\Phi_s} \exp\left(-\frac{1}{k_B T} U_S(\mathbf{r})\right)$$

where $\mathbf{R}_s$ is the position of bead $s$ and $I_s$ its type (A or B). The Edwards Hamiltonian for the Gaussian chain is given by $H^G = \frac{3 k_B T}{2 a^2} \sum_{s=2}^{N_p} (\mathbf{R}_s - \mathbf{R}_{s-1})^2$.

To account for hydrodynamic flows, the time evolution of the density fields $\rho_I(\mathbf{r})$ is described by convection-diffusion equations with local kinetic coupling coefficient

$$\frac{\partial \rho_I(\mathbf{r},t)}{\partial t} + \nabla \cdot (\rho_I(\mathbf{r},t) \mathbf{u}(\mathbf{r},t))$$
$$= M_I \nabla \cdot \rho_I(\mathbf{r},t) \nabla \mu_I(\mathbf{r},t) + \xi_I(\mathbf{r},t) \qquad (3)$$

Here, $\mu_I = \delta F / \delta \rho_I$ is the intrinsic chemical potential,

$$\mu_I = -U_I + \sum_J \int d\mathbf{r}\, d\mathbf{r}'\, \varepsilon_{IJ}(|\mathbf{r} - \mathbf{r}'|) \rho_J(\mathbf{r}')$$
$$+ \kappa_H \upsilon_I \left(\sum_J \upsilon_J \rho_J - 1\right) \qquad (4)$$

$M_I$ is the mobility coefficient of beads of type $I$, and $\xi_I$ is a stochastic noise satisfying the fluctuation-dissipation relation

$$\langle \xi_I(\mathbf{r},t) \rangle = 0 \qquad ,$$
$$\langle \xi_I(\mathbf{r},t) \xi_J(\mathbf{r}',t') \rangle \qquad (5)$$
$$= -2 k_B T \delta_{IJ} M_I \nabla \cdot \rho_I(\mathbf{r},t) \nabla \delta(\mathbf{r}-\mathbf{r}') \delta(t-t')$$

The velocity field $\mathbf{u}$ of fluid obeys the continuity equation and the fluctuating Navier-Stokes equations, which read

$$\partial_t (\rho u_\alpha) + \partial_\beta (\rho u_\alpha u_\beta) = \partial_\beta \sigma_{\alpha\beta} + \partial_\beta \sigma'_{\alpha\beta} + f_\alpha - \partial_\beta P_{\alpha\beta}. \qquad (6)$$

Here $\rho = \sum m_I \rho_I$ is the total local *mass* density of the fluid, $\sigma_{\alpha\beta}$ and $\sigma'_{\alpha\beta}$ are deterministic and stochastic stress tensors, and the last two terms, $(f_\alpha - \partial_\beta P_{\alpha\beta})$, summarize the local volume forces acting on the fluid. The deterministic stress tensor $\sigma_{\alpha\beta}$ is linearly proportional to the velocity gradient, $\sigma_{\alpha\beta} = \eta_{\alpha\beta\gamma\delta} \partial_\gamma u_\delta$ with $\eta_{\alpha\beta\gamma\delta} = \eta_s (\delta_{\alpha\gamma} \delta_{\beta\delta} + \delta_{\alpha\delta} \delta_{\beta\gamma}) + (\eta_b - \frac{2}{3} \eta_s) \delta_{\alpha\beta} \delta_{\gamma\delta}$, where $\eta_s$ is the shear viscosity and $\eta_b$ the bulk viscosity. The fluctuating stress $\sigma'_{\alpha\beta}$ is related to the viscosity by means of the fluctuation-dissipation theorem

$$\langle \sigma'_{\alpha\beta}(\mathbf{r},t) \rangle = 0,$$
$$\langle \sigma'_{\alpha\beta}(\mathbf{r},t) \sigma'_{\gamma\delta}(\mathbf{r}',t') \rangle = 2 \eta_{\alpha\beta\gamma\delta} k_B T \delta(\mathbf{r}-\mathbf{r}') \delta(t-t') \qquad .(7)$$

Finally, the local volume forces, $(f_\alpha - \partial_\beta P_{\alpha\beta})$, may result from a pressure gradient or from external fields. In a one-component fluid with small density variations, the pressure tensor can be approximated by $P_{\alpha\beta} = \text{const.} + \delta_{\alpha\beta} c_s^2 \rho$, where $c_s$ is the isothermal speed of sound. We note that the constant does not contribute to Eq. (6) and can thus be omitted. In our multicomponent system, the coupling of the Navier-Stokes equations with the free energy density functional (1) can be implemented either by introducing an additional contribution to the pressure term, or a thermodynamic volume force $f_\alpha$. These two approaches are equivalent. Here we choose the force approach, i.e., we introduce the volume force[32]

$$\mathbf{f}(\mathbf{r},t) = -\sum_I \rho_I(\mathbf{r},t) \nabla \mu_I(\mathbf{r},t) \qquad (8)$$

This completes the definition of the basic model which must now be solved numerically. In the following, we propose a hybrid algorithm that solves the fluid dynamics part (Eq. (5)) via a multiple-relaxation-time lattice Boltzmann (LB) method and the density functional part (Eq. (3)) via a finite difference scheme.

In the LB method, the space is discretized by a lattice, and each lattice site $\mathbf{r}$ carries a set of discrete velocity distribution functions $n_i(\mathbf{r},t)$, which describe the partial density of fluid particles with velocity $\mathbf{c}_i$ at time $t$. The total density $\rho(\mathbf{r},t)$ and momentum density $\mathbf{j}(\mathbf{r},t)$ are related to the moments of the distribution functions[74]

$$\rho(\mathbf{r},t) = \sum_i n_i(\mathbf{r},t) \qquad ,$$
$$\mathbf{j}(\mathbf{r},t) = \rho(\mathbf{r},t) \mathbf{u}(\mathbf{r},t) = \sum_i n_i(\mathbf{r},t) \mathbf{c}_i + \tfrac{1}{2} \mathbf{f}\, \Delta t. \qquad (9)$$

The sum runs over the discrete velocity space $\{\mathbf{c}_i\}$. In the following, we adopt the popular D3Q19 model,[75] which is based on a cubic lattice and whose velocity set contains the zero velocity and the vectors that connect each node with its nearest and next-nearest neighbors. Here the velocity unit is the "lattice velocity" $c=h/\Delta t$ with the lattice parameter $h$ and the time step $\Delta t$.

Within one time step $\Delta t$, the distribution functions $n_i$ evolve according to

$$n_i(\mathbf{r}+\mathbf{c}_i\Delta t, t+\Delta t) = n_i^*(\mathbf{r},t)$$
$$:= n_i(\mathbf{r},t) + \Delta_i[\mathbf{n}(\mathbf{r},t)] + \Delta_i^f + \Delta_i' \quad (10)$$

i.e., the distributions are first reshuffled from $n_i$ to $n_i^*$ (collision step) and then shifted to neighboring sites according to their respective velocities (streaming step). The collision step mimics particle scattering events in the fluid ($\Delta_i[\mathbf{n}(\mathbf{r},t)]$) as well as the effects of the volume forces ($\Delta_i^f$) and stress fluctuations ($\Delta_i'$). Specifically, the scattering contribution relaxes the distributions $n_j$ towards the equilibrium distributions $n_j^{eq}$ with a linear collision operator $L_{ij}$ to be specified below, $\Delta_i[\mathbf{n}(\mathbf{r},t)] = L_{ij}(n_j - n_j^{eq})$. In the D3Q19 model, the equilibrium distribution for given density $\rho$ and fluid velocity $\mathbf{u}$ is given by

$$n_i^{eq}(\rho,\mathbf{u}) = \rho\,\omega_i\left(1 + \frac{u_\alpha c_{i\alpha}}{c_s^2} + \frac{1}{2}\frac{u_\alpha u_\beta c_{i\alpha} c_{i\beta}}{c_s^4} - \frac{1}{2}\frac{u^2}{c_s^2}\right), \quad (11)$$

where $c_s = c/\sqrt{3}$ is the isothermal speed of sound, and the weight coefficients are $\omega_0 = 1/3$, $\omega_{1-6} = 1/18$, and $\omega_{7-18} = 1/36$. It is constructed as an expansion in the velocity up to second order that satisfies the constraints $\sum_i n_i^{eq} = \rho$, $\sum_i n_i^{eq} c_{i\alpha} = \rho u_\alpha$, and $\sum_i n_i^{eq} c_{i\alpha} c_{i\beta} = \rho u_\alpha u_\beta + P_{\alpha\beta}$ with $P_{\alpha\beta} = \delta_{\alpha\beta}\,c_s^2\rho$.

The simplest choice of the collision operator[18] is $L_{ij} = -\delta_{ij}/\tau$, corresponding to a single relaxation time that determines both the bulk and shear viscosity. In this paper, we employ a more sophisticated scheme, the multiple-relaxation-time model[39,40,41], which has the advantage that the mass and momentum are conserved rigorously and that both viscosities can be varied independently. The scheme has recently been reviewed in detail by Dünweg and coworkers[19,76] and we will refer the reader to these references for details and derivations. Basically, one constructs eigenvectors $\mathbf{e}_k$ of the collision operator $L_{ij}$ from outer products of the velocities $\mathbf{c}_i$, which are mutually orthogonal with respect to the scalar product $\sum_i \omega_i\, e_{ki} e_{kl}$ and relaxes them individually with relaxation parameters $\gamma_k$, corresponding to relaxation rates $(1-\gamma_k)/\Delta t$. The first four eigenvectors, $\mathbf{e}_0$-$\mathbf{e}_3$, correspond to the conserved modes (mass and momentum) and have the relaxation parameter $\gamma_{0-3}=1$. In choosing the higher eigenvectors, one has some freedom. The D3Q19 eigenvectors and relaxation parameters used in the present work are summarized in Table 1.

**Table 1** Basis set of the D3Q19 model
$\mathbf{e}_{ki}$ denotes eigenvector and $\gamma_k$ the relaxation parameter.

| $k$ | $\mathbf{e}_{ki}$ | $\gamma_k$ | $k$ | $\mathbf{e}_{ki}$ | $\gamma_k$ |
|---|---|---|---|---|---|
| 0 | 1 | 1 | 10 | $(3\mathbf{c}_i^2 - 5c^2)c_{ix}/c^3$ | 0 |
| 1 | $c_{ix}/c$ | 1 | 11 | $(3\mathbf{c}_i^2 - 5c^2)c_{iy}/c^3$ | 0 |
| 2 | $c_{iy}/c$ | 1 | 12 | $(3\mathbf{c}_i^2 - 5c^2)c_{iz}/c^3$ | 0 |
| 3 | $c_{iz}/c$ | 1 | 13 | $(c_{iy}^2 - c_{iz}^2)c_{ix}/c^3$ | 0 |
| 4 | $\mathbf{c}_i^2/c^2 - 1$ | $\gamma_b$ | 14 | $(c_{iz}^2 - c_{ix}^2)c_{iy}/c^3$ | 0 |
| 5 | $(c_{ix}^2 - c_{iy}^2)/c^2$ | $\gamma_s$ | 15 | $(c_{ix}^2 - c_{iy}^2)c_{iz}/c^3$ | 0 |
| 6 | $(\mathbf{c}_i^2 - 3c_{iz}^2)/c^2$ | $\gamma_s$ | 16 | $3\mathbf{c}_i^4/c^4 - 6\mathbf{c}_i^2/c^2 + 1$ | 0 |
| 7 | $c_{ix}c_{iy}/c^2$ | $\gamma_s$ | 17 | $(2\mathbf{c}_i^2 - 3c^2)(c_{ix}^2 - c_{iy}^2)/c^4$ | 0 |
| 8 | $c_{ix}c_{iz}/c^2$ | $\gamma_s$ | 18 | $(2\mathbf{c}_i^2 - 3c^2)(\mathbf{c}_i^2 - 3c_{iz}^2)/c^4$ | 0 |
| 9 | $c_{iy}c_{iz}/c^2$ | $\gamma_s$ | | | |

The nonequilibrium distribution $(n_i - n_i^{eq})$ and the forcing terms $\Delta_i^f$ are then expanded in moments using the basis vectors $e_{ki}$[19,76]

$$m_k = \sum_i (n_i - n_i^{eq})\, e_{ki}, \qquad m_k^f = \sum_i \Delta_i^f e_{ki}, \quad (12)$$

which evolve during the collision step according to

$$m_k^* = \gamma_k m_k + m_k^f + \varphi_k r_k. \quad (13)$$

Here $\varphi_k = \sqrt{\rho k_B T b_k (1-\gamma_k^2)/(c_s^2 h^3)}$ is the amplitude of the random noise for the $k$-th mode with $b_k = \sum_i \omega_i e_{ki}^2$, and $r_k$ is a Gaussian random number with zero mean and unit variance. Exploiting the orthogonality of the eigenvectors, one can easily reconstruct the new distribution function after collision, $n_i^* = n_i^{eq} + \omega_i \sum_k b_k^{-1} m_k^* e_{ki}$. Finally, the collision operators of the forcing terms are given by[19]

$$\Delta_i^f = \frac{\omega_i \Delta t}{c_s^2}\left[f_\alpha c_{i\alpha} + \frac{1}{2c_s^2}\Sigma_{\alpha\beta}(c_{i\alpha}c_{i\beta} - c_s^2\delta_{\alpha\beta})\right] \quad (14)$$

with

$$\Sigma_{\alpha\beta} = \frac{1}{2}(1+\gamma_s)\left(u_\alpha f_\beta + u_\beta f_\alpha - \tfrac{2}{3}u_\gamma f_\gamma \delta_{\alpha\beta}\right) + \frac{1}{3}(1+\gamma_b)u_\gamma f_\gamma \delta_{\alpha\beta}.$$

The multiple-time relaxation LB model described above solves Eq. (5) for a pressure term of the form $P_{\alpha\beta} = \text{const.} + \delta_{\alpha\beta} c_s^2 \rho$ and arbitrary volume forces **f**. The shear viscosity $\eta_s$ and bulk viscosity $\eta_b$ are related to $\gamma_s$ and $\gamma_b$, respectively, via[76]

$$\eta_s = \frac{\Delta t\, \rho c_s^2}{2} \frac{1+\gamma_s}{1-\gamma_s}, \qquad \eta_b = \frac{\Delta t\, \rho c_s^2}{3} \frac{1+\gamma_b}{1-\gamma_b} \quad (15)$$

The remaining equations, the set of convection-diffusion equations (3), are integrated numerically via a finite-difference scheme on the same lattice nodes used for the lattice Boltzmann method. The diffusion part of the equations has been implemented by the implicit Crank-Nicholson algorithm[77] and the convection term of the equations is discretized using the explicit two-step Lax-Wendroff scheme.[78] The thermal noise terms $\xi_I(\mathbf{r},t)$ are calculated by an algorithm due to Vlimmeren and Fraaije.[79] The Gaussian chain density functional of I-type bead is expressed in terms of Green's propagators

$$\rho_A(\mathbf{r}) \propto \sum_{s=1}^{n} G_s(\mathbf{r}) \sigma\left[G_{s+1}^+\right](\mathbf{r})$$

$$\rho_B(\mathbf{r}) \propto \sum_{s=n+1}^{N_P} G_s(\mathbf{r}) \sigma\left[G_{s+1}^+\right](\mathbf{r}), \quad (16)$$

where the set of integrated Green's functions $G_s$ and inverse Green's functions $G_s^+$ are related by recurrence relations

$$G_s(\mathbf{r}) = \exp\left(-\frac{U_{I_s}(\mathbf{r})}{k_B T}\right) \sigma[G_{s-1}](\mathbf{r})$$

$$G_s^+(\mathbf{r}) = \exp\left(-\frac{U_{I_s}(\mathbf{r})}{k_B T}\right) \sigma[G_{s+1}^+](\mathbf{r}) \quad (17)$$

subject to the initial conditions $G_0(\mathbf{r}) = G_{N_P+1}^+(\mathbf{r}) = 1$, and the linkage operator $\sigma$ denotes convolution with the Gaussian kernel

$$\sigma[f](\mathbf{r}) = \left(\frac{3}{2\pi a^2}\right)^{2/3} \int_V d\mathbf{r}' \exp\left(-\tfrac{3}{2a^2}(\mathbf{r}-\mathbf{r}')^2\right) f(\mathbf{r}').$$

This completes the description of our hybrid LB/SCF simulation method. In our simulations, we chose the time steps to be identical for the LB and the dynamic SCF simulations, i.e. LB steps alternate with propagations of the convection-diffusion equation (3). Multiple time step schemes are also conceivable, but have not been explored here. We note that every propagation step of Eq. (3) involves the self-consistent determination of the potentials $U_I$ (see Eq. (4)) from Eqs. (16) and (17). This was done with the Fletcher-Reeves nonlinear conjugate gradient method. The Gaussian integrals (17) were integrated with a 27-point stencil scheme[26].

In the following section, we will describe a number of applications to copolymer systems. The bead volume, the bead mass, and the mobility coefficient in these studies are taken to be equal for all bead species, i.e., $\upsilon_A = \upsilon_B = \upsilon_S =: \upsilon_0$, $m_A = m_B = m_S =: m_0$, and $M_A = M_B = M_S =: M$. The basic length unit is the lattice parameter $h$, the basic energy unit is the thermal energy $k_B T$, and the basic time unit is $\tau = h^2/k_B T\, M$. In the remaining paper, all quantities shall be given in these units. For example, the basic viscosity unit is $\eta^* = k_B T \tau / h^3$.

An important parameter of the simulation is the time step $\Delta t$, which sets the lattice velocity $c = h/\Delta t$ and hence the speed of sound $c_s$ in the model. In a typical simulation setup, $c_s$ is much smaller than the real speed of sound. This is not a problem since we are interested in slow diffusive processes and $c_s$ is not a physically relevant parameter in our context. However, we must carefully ensure that our processes are indeed slow compared to the value of $c_s$ in our model.

The other model parameters are the statistical segment length, which we set to $a = 1.1543\, h$ for practical reasons[80], the "bead" mass $m_0$, which we set to $m_0 = 1\, k_B T \tau^2/h^2$, the "bead" volume $\upsilon_0$, which sets the noise scaling parameter[81] $\Omega = h^3/\upsilon_0$, the Helfand compressibility parameter $k_H$, which we set to $k_H \upsilon_0/k_B T = 10$, the Flory-Huggins parameters $\chi_{IJ}^0 = \left(2\varepsilon_{IJ}^0 - \varepsilon_{II}^0 - \varepsilon_{JJ}^0\right)/(2k_B T \upsilon_0)$, and the bulk and shear viscosity $\eta_s$ and $\eta_b$, which are related to the relaxation parameters $\gamma_s$ and $\gamma_b$ via Eq. (15).

Unless stated otherwise (i.e., for all simulations except those belonging to Figure 2), we have chosen the noise parameter $\Omega = 400$ and our simulations were carried out in boxes of size $64h \times 64h \times 64h$ with period boundary conditions, using a time step of $\Delta t/\tau = 0.4$. For every system and every choice of parameter set, we have carried out five independent runs with different random seeds.

We can get a rough idea about the values of our simulation units in terms of SI units by matching the temperature T=300K, the radius of gyration of the polymers $R_g$, and their cooperative diffusion constant $D_P$. In the simulations, we have $R_g = a\sqrt{N/6}$ and $D_P \sim M/k_B T N$ in the Rouse regime, with simulated chains of length $N\sim10$ beads. Mapping real chains of size $R_g\sim3$nm, we conclude that our length unit corresponds to $h\sim2$nm. The diffusion constant of polymers is of order $D_p\sim10^{-6}$cm$^2$/s in solution, and $D_p\sim10^{-9}$cm$^2$/s in dense melts. Mapping this to the simulations yields the time units $\tau\sim4\cdot10^{-9}$s in solution, and $\tau\sim4\cdot10^{-6}$s in dense melts. The corresponding basic viscosity unit is $\eta^*\sim2\cdot10^{-3}$Ns/m$^2$ in solution, which is of the order of the shear viscosity of water ($\eta\sim0.8\cdot10^{-3}$Ns/m$^2$), and $\eta^*\sim2\cdot$Ns/m$^2$ in dense melts.

## 3. RESULTS

### 3.1 Dynamics of Microphase Separation in Block Copolymer Melts

We first use our method to investigate the effect of hydrodynamics on the kinetics of microphase separation in block copolymer melts. To this end, we have quenched systems of symmetric $A_8B_8$ block copolymers (Gaussian chains, $f=0.5$) from an initially disordered state into an ordered state with Flory-Huggins interaction parameter $\chi_{AB}^0 = 1.0$. This parameter has been chosen such that lamellar structures are generated according to the phase diagrams obtained by the static self-consistent field calculation.[82]. The initial configurations were thus homogeneous density distributions. In order to compare the results of our model with those of a system without hydrodynamic flows, we also performed simulations based on the diffusion-only model, which is defined by Eq. (3) with **u**=0.

Figure 1 shows the time evolution of three characteristic quantities during the microphase separation under different hydrodynamic conditions. The top panel displays the evolution with time of the free energy density and an order parameter defined as $S(t) = \sqrt{\frac{1}{V}\int_V \sum_I (\phi_I - \phi_I^0)^2}$, where $\phi_I = \rho_I v_0$ and $\phi_I^0$ are, respectively, the local and average volume fractions of I-type beads in the system. The lower panel shows two series of corresponding simulation snapshots, one for a purely diffusive system and one for a system with hydrodynamics (convection-diffusion model).

At the beginning, hydrodynamic flows are found to have little effect on the phase separation process (insets of Figure 1, snapshots at $t=200\tau$). A sharp increase of the order parameter in the initial stage of the simulation, accompanied by a rapid decrease of the free energy, indicates that the microphase separation sets in within $20\tau$ after the quench, and the curves for systems with and without hydrodynamics lie on top of each other. The simulation snapshots at times $200\tau$ illustrate that the patterns are still very similar. At this initial stage, hydrodynamic effects can thus be neglected.

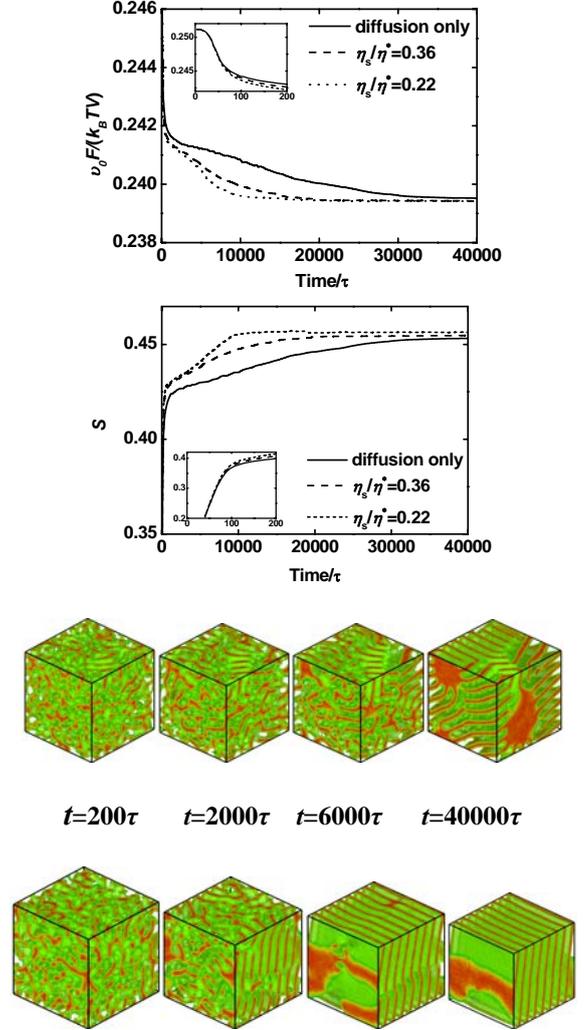

**Figure 1** Top panel: Time evolution of the free energy density $v_0 F/(k_B TV)$ and the segregation parameter $S$ during the microphase separation process of $A_8B_8$ block copolymers after a quench from the disordered state, with and without hydrodynamics for various values of shear viscosity at fixed bulk viscosity $\eta_b=0.34\eta^*$. The insets show blowup for early times. Error bars are omitted for clarity. Middle and bottom panel: Corresponding configuration snapshots without hydrodynamics (middle) and with hydrodynamics (bottom, $\eta_s=0.22\eta^*$ and $\eta_b=0.34\eta^*$) at times as indicated. Red regions indicate A-rich regions, green surfaces show isodensity surfaces $\phi_A=0.5$.

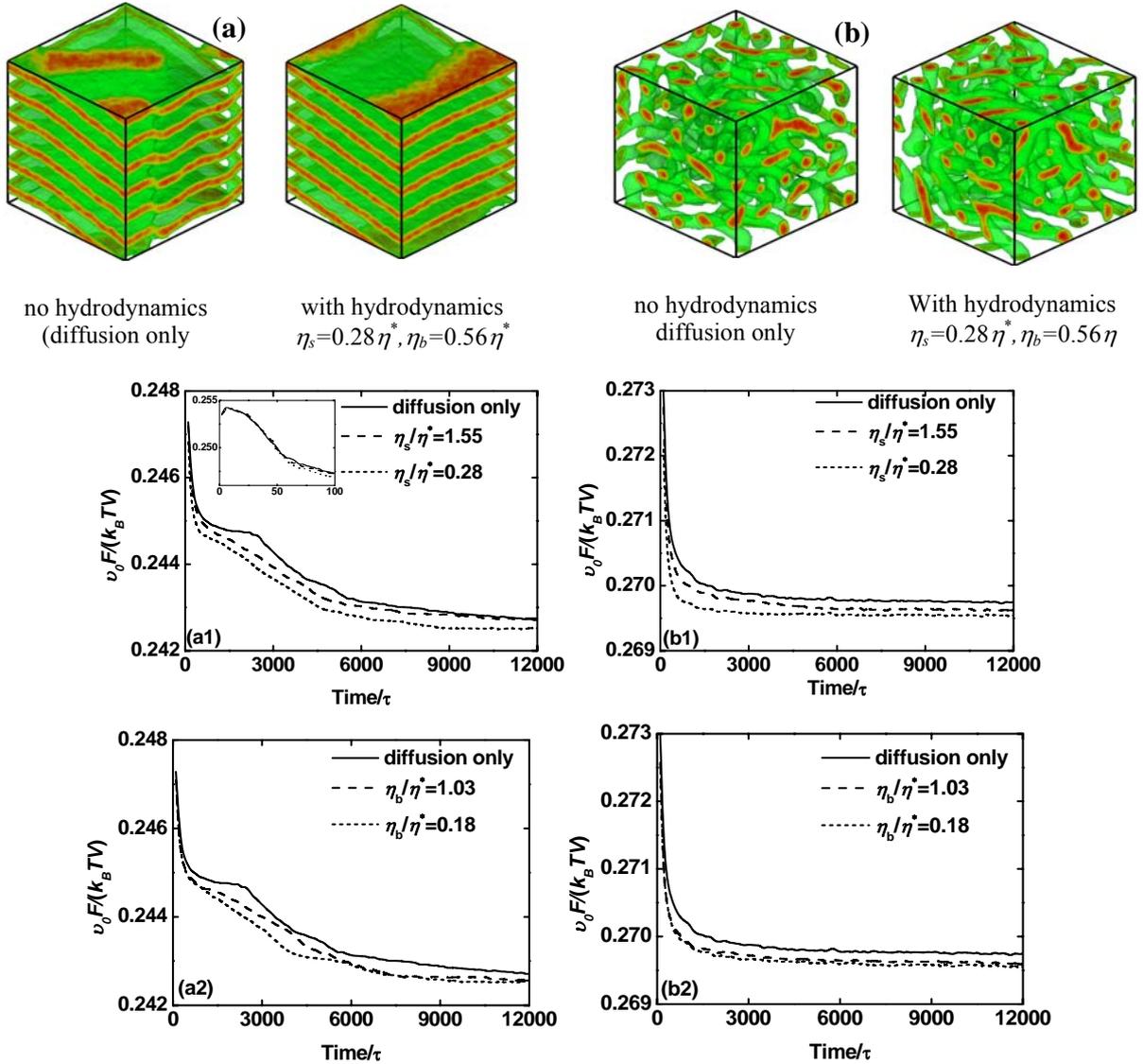

**Figure 2** Microphase separation in a system of (a) symmetric ($A_8B_8$) and (b) asymmetric ($A_5B_{11}$) block copolymers upon quenching into the ordered phase from an initially disordered solution, without and with hydrodynamics. Top: Final configuration after $12000\,\tau$ without and with hydrodynamics (color coding as in Figure 1). Bottom: Evolution of free energy (a1, b1) for various values of shear viscosity $\eta_s$ at fixed bulk viscosity $\eta_b=0.56\eta^*$; (a2, b2) various values of $\eta_b$ at fixed $\eta_s=0.83\eta^*$. Error bars are omitted for clarity. Inset in (a1) shows blowup in the initial stage. In all simulations, the noise parameter for the convection-diffusion dynamics was set to $\Omega=100$ and the hydrodynamic noise was turned off.

Shortly after that, however, the hydrodynamic flows begin to have a pronounced effect on the time evolution. In the presence of hydrodynamic interactions, the system orders faster, the free energy drops faster, and the ordered domains grow faster. Both the diffusive system and the systems with hydrodynamics tend to get trapped in metastable states directly after the quench. This leads to the formation of a plateau in the free energy curve at early times, $t\sim 1000\,\tau$. In the presence of hydrodynamics, the systems manage to escape this metastable plateau within the time $t\sim 10000\,\tau$. Purely diffusive systems remain trapped and order only very slowly. Even though they show local lamellar order at the end of the simulation runs ($40000\,\tau$), the configurations are still filled with defects. The mean free energy is still higher than that of systems with hydrodynamic interactions. Thus the main effect of convective flows on the evolution process of the microstructures is to promote the removal of defects. The lower the shear viscosity, the more effective they are. We conclude that hydrodynamic interactions help the system to overcome free energy barriers and to access its final equilibrium ordered state.

Next we have studied the influence of different model parameters on this effect, i.e., we have varied the bulk and shear viscosity separately, and we have also considered asymmetric copolymers that order into a different morphology. To assess the influence of the noise, we have also chosen a different noise scaling parameter, $\Omega=100$, and turned off the noise in the LB part of the simulations (i.e., $\varphi_k=0$ in Eq. (13), which corresponds to setting $\sigma'_{\alpha\beta}=0$ in Eq. (6)). These studies were carried out in smaller systems of size $48h \times 48h \times 48h$ and a time step $\Delta t/\tau=0.2$. The results are summarized in Figure 2.

Figure 2(a) shows the evolution of systems of symmetric $A_8B_8$ block copolymers after quenching from the disordered phase into the lamellar phase at $\chi^0_{AB}=1.0$ for various combinations of shear and bulk viscosity. In Figure 2 (a1), the bulk viscosity $\eta_b$ is kept constant and the shear viscosity $\eta_s$ is varied, which corresponds to the situation studied in Figure 1. The results are very similar. Hydrodynamic flows accelerate the ordering and help to remove lamellar defects. The ordering in the diffusive case is faster than in the systems of Figure 1, which can presumably be attributed to the smaller system size and/or the higher noise level. In the presence of the long-range hydrodynamic interactions, the ordering time becomes comparable. Furthermore, the free energy density of the equilibrated final state is higher than in Figure 1. This is also an effect of the higher noise level (the amplitude of the noise is inversely proportional to $\sqrt{\Omega}$) and the ensuing slightly weaker segregation of the components. Indeed, the order parameter in the system of Figure 2(a) tends to $S \rightarrow 0.425$ at large times (data not shown) as opposed to $S \rightarrow 0.46$ in Figure 1. Apart from this slight effect on the final equilibrium state, the noise has no qualitative influence on the results. Hydrodynamic interactions are still found to accelerate the ordering during the whole process of microphase separation, until equilibrium is reached.

This is different for asymmetric block copolymers. Figure 2(b) shows the time evolution of the free energy density during the microphase separation of $A_5B_{11}$ block copolymers ($f=0.31$) for the cases of the diffusion-only model and different realizations of the convection-diffusion model. The systems were quenched from an initially disordered state $\chi^0_{AB}=0$ to an ordered state at $\chi^0_{AB}=1.3$. Under these conditions, the asymmetric block copolymers should self-assemble into a hexagonal array of cylindrical microstructures according to the mean-field theory.[82] Initially, the behaviour of the free energy density is similar to that of symmetric block copolymers. It decreases more rapidly in the presence of convective flows than in purely diffusive systems. The hydrodynamic interactions promote the ordering process at early times. Once the system is locally ordered, however, they are no longer effective, and the system remains trapped in metastable states just like in the diffusive case. The final snapshots after the total run length of $12000\tau$ feature a disordered mess of entangled cylindrical microstructures, which looks very much alike for the diffusive and the convective-diffusive system. In contrast, the convective flows in systems of symmetric block copolymers help to remove the defects of microstructures not only in the early stage, but also at intermediate and late stages of the microphase separation.

Inspecting in more detail the influence of the different viscosity parameters, we finds that the key quantity is clearly the shear viscosity $\eta_s$. Varying the bulk viscosity $\eta_b$ at fixed shear viscosity has only a small effect on the time evolution (Figure 2 a2 and b2). By varying the shear viscosity $\eta_s$, on the other hand, one can influence the ordering process significantly (Figure 1 and Figure 2, a1 and b1). The hydrodynamic interactions are most effective at small shear viscosity $\eta_s$.

To summarize this subsection, we have applied our new method to study microphase separation in block copolymer melts as a first test case. We found, in agreement with earlier simulations by Maurits et al[32], that hydrodynamics do not matter in the initial stage of microphase separation, but that they accelerate the ordering in the later stages. In systems of symmetric block copolymers that order into lamellar phases, hydrodynamic interactions are found to be very effective in removing local defects throughout the whole ordering process. This is no longer true for asymmetric block copolymers that order into hexagonal structures. In that case, the systems remain trapped in configurations full of defects even in the presence of hydrodynamic flows. Hence the effect of hydrodynamic interactions is found to be less pronounced for asymmetric copolymers than for symmetric copolymers, in agreement with the earlier cell-dynamics simulations by Yokojima and Shiwa[11]. Furthermore, we find the shear viscosity to be much more important than the bulk viscosity, in agreement with theoretical expectations in the Stokes limit, where the quantity driving the long-range hydrodynamic interactions is the shear viscosity, and the bulk viscosity plays only a minor role. Thus our method applied to block copolymers gives reasonable results compared to earlier work based on different methods as well as to theoretical expectations. This encourages us to apply it to a more complex system in the next step.

## 3.2 Kinetics of Vesicle Formation of Amphiphilic Block Copolymers in Solution

As a second application example, we study the kinetics of nanostructure self-assembly in copolymer solutions. The model block copolymers are represented by $A_2B_8$ Gaussian chains in a solvent S. The Flory-Huggins interaction parameters are set to $\chi^0_{AB}=2.3$, $\chi^0_{BS}=2.4$, and $\chi^0_{AS}=0.25$. Thus A blocks and B blocks are incompatible, the A block is hydrophilic, and the B block is hydrophobic. The average volume fraction of amphiphilic block copolymers in solution is set to 0.2. These parameters are chosen in the spinodal region of the homogenous mean-field phase diagram[83], i.e., the system will phase separate after quenching from the homogenous state[69]. Mapping the diffusion constant and the radius of gyration of the molecules to those of real copolymer systems[71], we find that our length unit, $h$, is in the range of a few nm, and the time unit $\tau$ corresponds to roughly $10^{-9}$s.

In this work, we focus on the effect of hydrodynamic flows on the dynamics of spontaneous vesicle formation and vesicle fusion. Motivated by the previous finding in Section 3.1 that the bulk viscosity $\eta_b$ is not a very influential parameter, we fix $\eta_b/\eta^*=0.34$ and vary the value of the shear viscosity $\eta_s$ only. Like before, we have also performed based on the diffusion-only model for comparison.

Figure 3 compares the evolution with time of the free energy density and the degree of segregation, characterized by the quantity $S=\frac{1}{V}\int d\mathbf{r}|\phi_A(\mathbf{r})+\phi_B(\mathbf{r})-\phi^0_A(\mathbf{r})-\phi^0_B(\mathbf{r})|$, during the aggregation process in the absence and presence of hydrodynamic flows. Like in the copolymer melt, the effect of hydrodynamic interactions is small in the initial stage (inset). The convective flows have only a weak effect on the spinodal decomposition of the amphiphilic block copolymers. After an initial "incubation time" where spinodal fluctuations build up until they reach a critical threshold[69], the free energy suddenly drops and the segregation parameter rises sharply, indicating that the amphiphilic block copolymers begin to aggregate into clusters. This is followed by a period of cluster growth and reorgani-zation, which is accompanied by slower further segregation and free energy decrease. In these intermediate and late stages, hydrodynamic flows become important and significantly accelerate the self-organization process. Like in the case of copolymer melts, the hydrodynamic effects on the free energy are strongest for small shear viscosities.

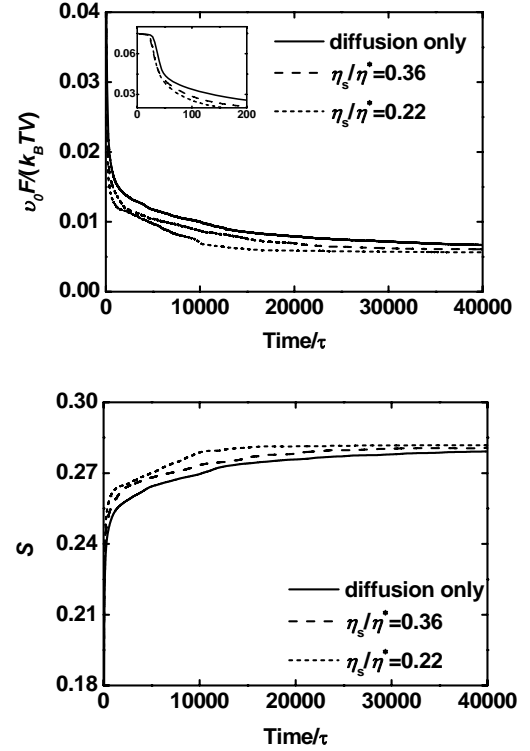

**Figure 3** Time evolution of the free energy density $\upsilon_0 F/(k_B TV)$ and the segregation parameter during the quenching process of amphiphilic block copolymers in solution for simulations with and without hydrodynamics. The inset shows the free energy density as a function of time in the initial stage. The bulk viscosity is $\eta_b=0.34\eta^*$, the shear viscosity $\eta_s$ is varied as indicated in the graphs.

The corresponding simulation snapshots provide direct insight into the self-assembly behavior of the amphiphilic block copolymers. Figure 4 shows the evolution of the aggregate morphologies of amphiphilic block copolymers in solution within the diffusion-only model (4a) and the convection-diffusion model (4b). After having been quenched from the initially homogeneous state, the amphiphilic block copolymers rapidly aggregate into a large number of spherical micelles. These micelles then merge with neighbor micelles and grow (snapshots a2 and b2), until they reach a critical size where the micelle core can no longer accommodate the chains. At this stage, the micelles reorganize themselves into semivesicles which contain A-blocks in the center (snapshots a3 and b2). The semi-vesicles further coalesce with neighboring small spherical micelles, solvent diffuses into the inner parts of aggregates until, finally, vesicles are formed (snapshots a4 and b4). In the late stage of the process, vesicle fusion events are also observed, which is why some of the vesicles in Figures 4, a4 and b4 are elongated.

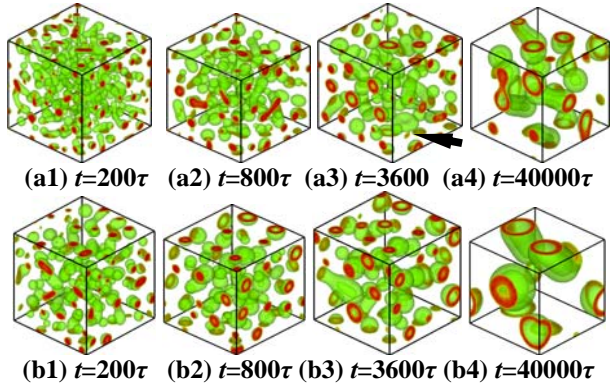

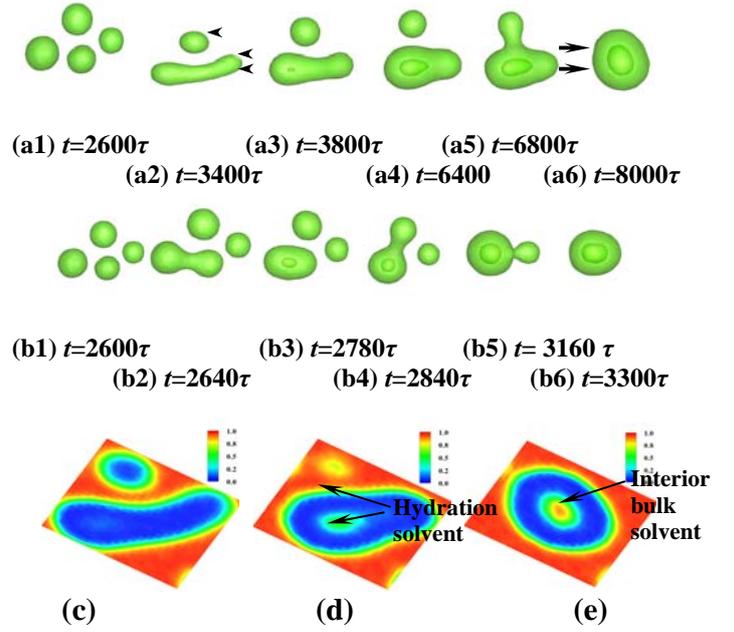

**Figure 4** Pattern evolution during the self-assembly of amphiphilic block copolymers (a) in the diffusion-only model (top) and (b) with hydrodynamics (bottom, shear viscosity $\eta_s=0.22\eta^*$, and bulk viscosity $\eta_b=0.34\eta^*$) at simulation times $t$ as indicated. Colour coding as in Fig.1. The arrow in (a3) points to a bilayer.

The self-assembly of vesicles from micelles via semi-vesicles is illustrated in more detail in Figure 5. Here, we start from an identical configuration containing several micelles, which has been generated by a diffusion-only simulation, and monitor its evolution in the presence and absence of hydrodynamic flows. The pathway of vesicle formation is similar in both cases. First, spherical micelles coalesce to form a cylindrical micelle (snapshots a2 and b2). The solvent content in the inner parts of the micelle is very small due to the hydrophobicity of the B-beads (Figure 5c). Then the short cylindrical micelle gradually retracts towards a spherical shape. Once the micelle diameter exceeds a certain size in the process of retraction, the amphiphilic block copolymers start to flip-flop and reorganize into a semi-vesicle (snapshots a3-a4, b3-b4). The solvent diffuses through the hydrophobic B-bead regions and swells the hydrophilic A-beads in the cores of aggregates. Thus the semi-vesicle contain some hydration solvent in the core part, but no interior bulk solvent (Figure 5d). The semi-vesicle further merges with neighbour micelles (snapshots a5 and b5). The joint aggregate retracts again into a spherical shape and solvent diffuses into the inner part of the aggregate to form the interior bulk solvent (snapshots a6 and b6). The final vesicle contains a significant amount of interior bulk solvent (Figure 5e).

**Figure 5** Simulation snapshots during the late stage of vesicle formation in the (a) absence and (b) presence of hydrodynamic effects. The snapshots are extracted from identical regions of a much larger simulation box. The initial configuration at time $2600\tau$ is the same in both cases and was generated from a diffusion-only simulation. The convective flows were then turned on at $t=2600\tau$ in the simulation corresponding to in panel (b). In panels (a) and (b), the evolution time is shown below each snapshot. The shear viscosity and bulk viscosity are $\eta_s=0.22\eta^*$ and $\eta_b=0.34\eta^*$, respectively. Panels (c), (d), and (e) show the volume fraction distributions of solvent beads in a cross section of the aggregates marked with arrows in the panels (a2), (a4), and (a6) at time $3400\tau$, $6400\tau$, and $8000\tau$, respectively.

Comparing the simulations with and without hydrodynamic interactions, we find the pathways of vesicle formation to be qualitatively very similar. The sequence of structures, spherical micelle–cylindrical micelle–semi-vesicle–vesicle, is the same. However, the presence of hydrodynamic effects remarkably accelerates the formation of vesicles. For instance, at time $3300\tau$, the initial four spherical micelles have completed the whole process of evolution into a vesicle as the convective flows are included in the simulation (Figure 5, b6), whereas they have barely managed to merge into a short cylindrical micelle in the diffusion-only model (Figure 5, a2) and need another $4500\tau$ to reach the final vesicle state.

The pathway described above is the main mechanism of vesicle formation for the set of parameters chosen in this study. It is very similar to a pathway recently identified by He and Schmid [69,71]

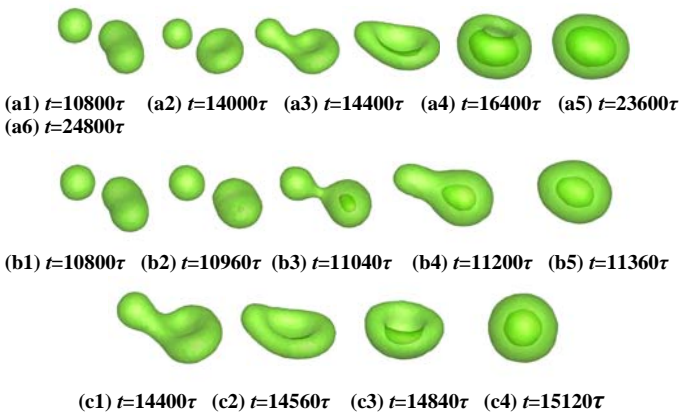

(a1) $t=10800\tau$ (a2) $t=14000\tau$ (a3) $t=14400\tau$ (a4) $t=16400\tau$ (a5) $t=23600\tau$
(a6) $t=24800\tau$

(b1) $t=10800\tau$ (b2) $t=10960\tau$ (b3) $t=11040\tau$ (b4) $t=11200\tau$ (b5) $t=11360\tau$

(c1) $t=14400\tau$ (c2) $t=14560\tau$ (c3) $t=14840\tau$ (c4) $t=15120\tau$

**Figure 6** (a) Representative snapshots during the process of vesicle formation via a platelet intermediate in the absence of hydrodynamics. (b) Snapshots during the vesicle formation in the presence of hydrodynamics. The convective flows have been turned on at time $10800\tau$. (c) Same as (b), but the convective flows are turned on at time $14400\tau$. The shear viscosity and bulk viscosity are $\eta_s=0.22\eta^*$ and $\eta_b=0.34\eta^*$, respectively.

from diffusion-only simulations. He et al used external potential dynamics[84], which is a dynamic SCF model with a non-local kinetic coupling coefficient that accounts for chain connectivity at the Rouse level. Like in the present study, the pathway of He et al. involves micelle growth at early times and a semi-vesicle intermediate state. However, coalescence and fusion events were much less prominent, i.e., the micelle and vesicle growth were mainly driven by attracting single copolymers from solution. This difference can partly be attributed to the different kinetic model (as will be discussed further below), and partly to the higher copolymer concentration in the present study. Indeed, when looking at higher copolymer concentrations in their model, He et al found a different pathway that involved micelle coalescence[71]: The micelles aggregated into flat bilayer disks which then curved around and closed up to form vesicles. The same mechanism had been observed in previous simulations by other authors [57,58,59] and is also occasionally found here in the diffusion-only simulations (an example of a bilayer intermediate is marked by an arrow in Figure 4, a3). It is not observed in the simulations with convective flow.

To elucidate the influence of hydrodynamic flows on this "disk pathway", we have carried out a set of simulations where convective flows are turned on at different stages of the vesicle formation process. The results are shown in Figure 6. The upper row (Figure 6a) shows the reference simulation, simulated within the diffusion-only model. At early stages, spherical micelles coalesce to form cylindrical micelles as above. The cylindrical micelle subsequently transforms itself into a flat bilayer platelet. The platelet merges with neighbour micelles and grows further into a curved disk. Then, the solvent is gradually encapsulated to reduce the energy contribution from the line tension of the rim. This results in a cuplike vesicle with a small pore. Eventually, the pore closes up and a vesicle is formed.

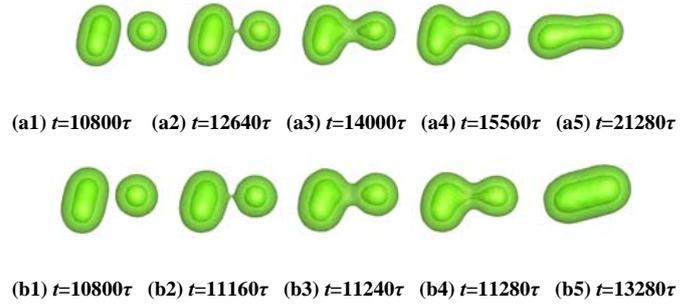

(a1) $t=10800\tau$ (a2) $t=12640\tau$ (a3) $t=14000\tau$ (a4) $t=15560\tau$ (a5) $t=21280\tau$

(b1) $t=10800\tau$ (b2) $t=11160\tau$ (b3) $t=11240\tau$ (b4) $t=11280\tau$ (b5) $t=13280\tau$

**Figure 7** Representative snapshots during the fusion process of two small vesicles (a) in the absence and (b) presence of hydrodynamic effects. In panel (a), the convective flows are switched off at time $10800\tau$.

Starting from this diffusion-only simulation, we have investigated the effect of switching on convective flows at different times. The results depend crucially on whether this happens before or after the bilayer disks have formed. In the first case, the disk formation is prevented and the vesicle formation proceeds via the semi-vesicle pathway (Figure 6b). In the second case, the pathway of vesicle formation is similar to that found without hydrodynamic flows, but the rate of pore closure is remarkably sped up by the convective flows (Figure 6c). Thus we conclude that the convective flows not only accelerate the formation of vesicles, but also contribute to selecting the pathway of vesicle formation, and that they favor the semi-vesicle pathway.

Next we discuss the effect of hydrodynamics on the fusion of vesicles. Figure 7 shows details of a fusion process of vesicles in the absence and presence of convective flows. The starting configuration has been obtained with the convection-diffusion model. Then two parallel simulations were carried out, one with and one without convective flows. Figure 7 illustrates that the whole process of vesicle fusion in our system is in line with the classic stake-pore fusion mechanism.[85,86,87] Initially, at time 10800$\tau$, the two small vesicles are separated. Then the amphiphilic block copolymers in the outer layers of the vesicles rearrange locally and bridge the solvent gap between the vesicles. This results in the formation of a stalk (snapshots a2 and b2). Subsequently, the stalk intermediate expands and the outer layers of vesicles recede. The inner layers of vesicles merge and produce a hemifusion diaphragm (snapshots a3 and b3). Finally a fusion pore appears (snapshots a4 and b4). At this time, the fusion process of vesicles is completed. The freshly formed vesicle retracts toward a spherical shape, but the rate of retraction is very slow (snapshots a5 and b5). The same mechanism is observed both in the absence and presence of convective flows (Figure 7a and b). However, the long-range hydrodynamic interactions accelerate the process by an order of magnitude. The total time of vesicle fusion takes about 2920$\tau$ in the diffusion-only model (from snapshot a2 to a4), and 120$\tau$ in the presence of hydrodynamics (from snapshot b2 to b4).

We should note that fusion events were much more frequent in the present study than in the previous studies by He et al[70,71,72], even in the diffusive case. This is most likely an artefact of the local dynamic SCF approach chosen in the present study. As mentioned above, He et al used external potential dynamics[84], which is a dynamic SCF theory with nonlocal mobility coefficient that accounts for the chain connectivity. Fusion events practically never occurred in the later stages of the simulations once the vesicles were fully developed. This is consistent with experiments where spontaneous fusion events are practically never observed. In contrast, Sevink and Zvelindovsky[67,68] observed fusion and aggregation in their studies of compact copolymer vesicles in solution, much like in the present study. They also used a local version of dynamic SCF theory.

To conclude our analysis of hydrodynamic effects on vesicle formation, we will now examine a number of quantities that characterize the vesicle configurations at a global level. One highly relevant quantity is the solvent content inside the vesicles. Figure 8 shows the volume fraction of trapped solvent as a function of time during the evolution process of the aggregates. The trapped solvent is confined in the inner layers of vesicles and consists of the hydration solvent of the inner leaflets and the interior bulk solvent, which were illustrated in Figures 5d and 5e, respectively. The volume fraction of trapped solvent is given by $f_{trap}=V_{trap}/V_{tot}$, where $V_{trap}$ and $V_{tot}$ are the volume of the trapped solvent and the total solvent in the system, respectively. In the early stages of aggregation, where the amphiphilic block copolymers are still mainly organized in micelles and semi-vesicles, the amount of trapped solvent is small. At this stage, the error bars are quite small, indicating that there is little statistical spread in the amount of trapped solvent: As long as most of the trapped solvent is hydration solvent, its volume fraction is closely linked to the distribution of A and B blocks in the aggregates and there is little room for variations. This changes as soon as the semi-vesicles begin to fill with interior bulk solvent, which happens around $f_{trap}\sim0.006$. The error bars increase dramatically and the amount of trapped solvent rises, first linearly up to $f_{trap}\sim0.06$, and then slowly saturating. In the presence of hydrodynamic flows, this process sets in earlier and proceeds considerably faster than in the absence of hydrodynamics.

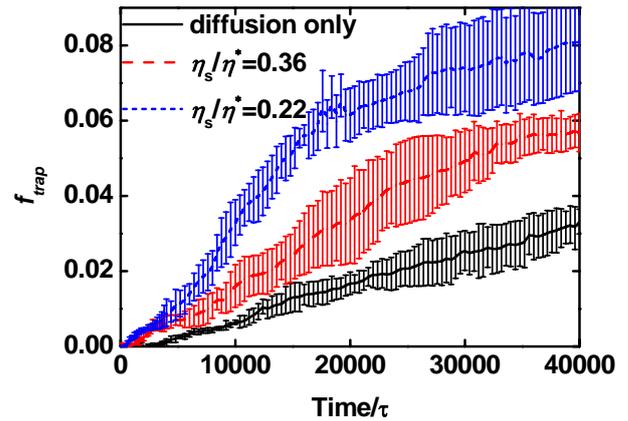

**Figure 8** Total volume fraction of trapped solvent as a function of time for the self-assembly of amphiphilic block copolymers. In the calculations, the trapped solvent consists of the interior bulk solvent and hydration solvent in the inner layers of vesicles, as illustrated in Figures 5e and 5d.

Further quantities characterizing the aggregate morphology are provided by the integral-geometry morphological measures, i.e., the Minkowski functionals. We refer to the reviews by Mecke for a detailed discussion[88,89]. The functionals in three dimensions include four independent geometrical quantities: The volume ($V$), the area ($A$), the mean curvature ($H$), and the Euler characteristic ($\chi_E$). The last quantity is related to the average Gaussian curvature and characterizes the topology. For instance, the Euler characteristic of spheres has a value of 2, and that of hollow spheres has a value of 4. The Euler characteristic of systems containing disjoint objects is equal to the sum of Euler characteristic of the individual objects. If two objects are linked by a passage, the Euler characteristic changes by (-2).

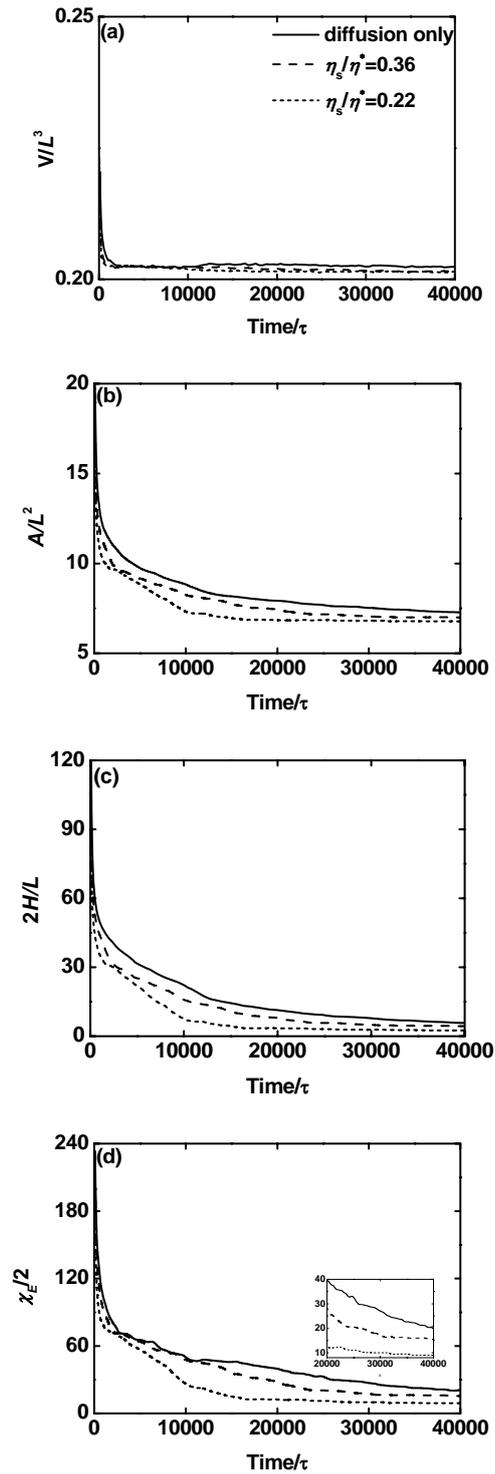

**Figure 9** Temporal evolutions of the Minkowski functionals for the hydrophobic B-beads during the quenching process of amphiphilic block copolymers with and without hydrodynamic effects. Panels (a), (b), (c), and (d) respectively show the volume ($V$), area ($A$), integral mean curvature ($H$), and Euler characteristic ($\chi_E$). $L$ is the size of simulation box. The inset of panel (d) shows the Euler characteristic as a function of time in the late stage.

Figure 9 displays the time dependence of the Minkowski functionals of the hydrophobic B-beads for the simulations with and without hydrodynamic effects. They have been calculated by a simple algorithm described by Michielsen and De Raedt,[90,91] with threshold value set to 0.20.[92] The evolution of the volume parameter is not very revealing. It sharply decreases in the early stage and then remains essentially constant throughout the simulations (Figure 9a). The area in Figure 9b also decreases in the early stage, which indicates that the micelles have a positive surface tension, and reaches a plateau value in the late stage, reflecting the fact that the total amount of interface in the equilibrium system is determined by the copolymer content. Similarly, the mean curvature drops during the evolution process, and then very slowly approaches a constant in the late stage (Figure 9c). This behaviour is characteristic for vesicles with positive bending stiffness. The Euler characteristic shows a more complex behaviour (Figure 9d). In the early stage, many spherical micelles are formed in the system and then quickly merge. This causes a rapid decrease of the number of aggregates, and the Euler characteristic decreases accordingly. In the following, the merging rate goes downs due to the exhaustion of small clusters, and at the same time micelles reorganize into semi-vesicles, which are topologically equivalent to hollow spheres. Due to the joint effect of these two processes, of which one reduces the Euler characteristics (the micelle coalescence), and one increases it (the transformation into semi-vesicle), the Euler characteristic almost stagnates for some time. In the late stage, the evolution of the system is dominated by vesicle fusion events and the Euler characteristic decreases again, as shown in the inset of Figure 9d. Comparing the evolution of the surface area, the mean curvature, and the Euler characteristic in the diffusion-only model and the convection-diffusion model for different shear viscosities, we find that the curves of course reflect the acceleration of the vesicle formation due to convective flows as expected. Otherwise, they are qualitatively similar. Thus hydrodynamic flows do not have an evident qualitative effect on the global morphological evolution of the system.

## 4. DISCUSSION AND CONCLUSION

To summarize, we have introduced a new method to treat hydrodynamic interactions in dynamic SCF simulations within a hybrid Lattice-Boltzmann/finite difference scheme, which solves a coupled set of SCF-based convection-diffusion equations and Navier-Stokes equations.

A few comments on the implementation of the method are in order. First, we have taken care to include fluctuations both in the convection-diffusion equations *and* in the Navier-Stokes equations. Thus we have implemented the fluctuating LB method, even though it is much less widely applied and computationally more expensive than regular, deterministic LB. The reason is that the amplitudes of the fluctuations in the convection-diffusion equations and the Navier-Stokes equations are intimately related. The noise level is basically given by the inverse of $\Omega = h^3/\upsilon_0$, which is the number of beads per lattice voxel. The smaller the number of beads per lattice point, the larger are the stochastic forces *both* in the discretized versions of the SCF equations and the fluctuating LB equations. In a consistent treatment, one must therefore either include them everywhere or omit them everywhere. In practice, however, we found that the treatment of the noise does not seem to be very critical. For comparison, we have also run the simulations of section 3.2 using the deterministic LB scheme (with noise switched off). The results were virtually identical to those presented here (data not shown). In most practical situations, it is thus probably acceptable to neglect the hydrodynamic noise. Nevertheless, we believe that such an approach is problematic from a conceptual point of view, because different degrees of freedom are coupled to heat baths of different temperature (T ~300 K for the composition degrees of freedom, T=0K for the flow degrees of freedom). This might lead to a constant unphysical heat flux from one thermal reservoir to the other and result in simulation artefacts, and even "violations" of the second law of thermodynamics.

Second, the dynamical model in our implementation is purely local in space and time and does not account for the effects of chain connectivity. For example, the monomer diffusion in Eq. (3) is driven by a local Onsager mobility coefficient. We have discussed in section 3.2 that the chain connectivity may influence the dynamics, e.g., it creates a kinetic barrier for fusion events. To overcome this problem, one can replace Eq. (3) by a convection-diffusion model based on "external potential dynamics"[84], which implements the dynamical behavior of Rouse chains. One subtle issue here is to implement the noise both efficiently and in a thermodynamically consistent manner, which can be done, but requires slight (well-motivated) changes in the underlying SCF free energy functional [29,93].

A more serious drawback of the present method is that it does not correctly account for the viscoelastic behaviour of dense polymers melts, unless the viscoelacity results from the microphase separation itself.[94,95] The hydrodynamic constitutive equations, $\sigma_{\alpha\beta} = \eta_{\alpha\beta\gamma\delta}\partial_\gamma u_\delta$, describe Newtonian fluids without memory. Furthermore, the SCF free energy functional is based on a reference ensemble of equilibrium polymers and does not include the possibility of chain stretching in shear flow. Our method can be refined in both respects. On the one hand, several approaches have been proposed to incorporate viscoelastic constitutive equations in LB simulations.[96,97,98,99,100] On the other hand, Maurits et al[101] have devised a heuristic method to incorporate chain stretching in shear flow by adjusting the reference ensemble of polymer configurations in the free energy functional. However, it is not clear how these different extensions should be combined in a unified and consistent way. Ideally, the macroscopic viscoelastic constitutive equation should emerge from a microscopic simulation model that accounts for chain stretching and the effect of entanglements on the diffusive motion of chains. The construction of such a model is far from trivial.

Nevertheless, the method presented here provides a powerful tool for studying unentangled polymer mixtures at low Weissenberg numbers and inhomogeneous polymer solutions. It can easily be extended to, e.g., multiphase fluids with locally varying or composition dependent viscosities or mass densities. In the present paper, we have applied it to different problems involving copolymer self-assembly, namely, microphase separation in diblock copolymer melts and vesicle formation in solutions of amphiphilic copolymers. Except in the initial stage of demixing, hydrodynamic flows were found to accelerate the process of self-assembly and ordering considerably, and also to influence the selection of pathways across the rugged free energy landscapes.

In the case of the vesicles, for example, they seem to act in favour of intermediates with spherical symmetry for the model parameters chosen in the present study. Future studies will be devoted to elucidate in more detail the mechanisms that determine the pathway selection – whether the crucial factor is the free energy of the intermediate structures, or whether they are dominated by kinetic factors.

## ACKNOWLEDGMENTS

We thank Andrei Zvelindovsky, and Burkhard Dünweg for helpful discussions. This work was funded by the VW foundation and the German Science Foundation (CRC TR6).

**Table of Contents Graphic**

**Hybrid Lattice Boltzmann/Dynamic Self-Consistent Field Simulations of Microphase Separation and Vesicle Formation in Block Copolymer Systems**

Liangshun Zhang* and Friederike Schmid

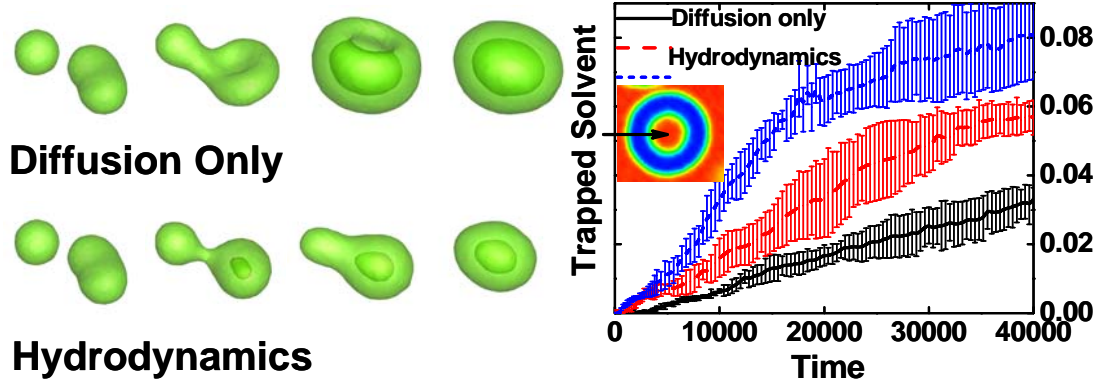